\documentclass[useAMS,usenatbib,usegraphicx]{mn2e}
\usepackage{pslatex}

\title[Extreme O and H in a Polluted White Dwarf]{Solar Abundances of Rock Forming Elements, Extreme Oxygen and Hydrogen in a 
Young Polluted White Dwarf}

\author[J. Farihi et al.]{J. Farihi$^{1}$\thanks{E-mail: j.farihi@ucl.ac.uk}\thanks{STFC Ernest Rutherford Fellow},
D. Koester$^2$,
B. Zuckerman$^3$,
L. Vican$^3$,
B. T. G\"ansicke$^4$,
N. Smith$^5$,
\newauthor
G. Walth$^5$,
E. Breedt$^4$
\\
$^1$Department of Physics and Astronomy, University College London, London WC1E 6BT\\
$^2$Institut f\"ur Theoretische Physik und Astrophysik, University of Kiel, 24098 Kiel, Germany\\
$^3$Department of Physics and Astronomy, University of California, Los Angeles CA 90095, USA\\
$^4$Department of Physics, University of Warwick, Coventry CV4 7AL\\
$^5$Steward Observatory, University of Arizona, Tucson AZ 85721, USA}

\begin{document}

\date{}

\maketitle

\label{firstpage}

\begin{abstract}
The $T_{\rm eff}=20\,800$\,K white dwarf WD\,1536$+$520 is shown to have broadly solar abundances of the major rock forming elements 
O, Mg, Al, Si, Ca, and Fe, together with a strong relative depletion in the volatile elements C and S.  In addition to the highest metal abundances 
observed to date, including $\log$\,(O/He) $=-3.4$, the helium-dominated atmosphere has an exceptional hydrogen abundance at $\log$\,(H/He) 
$=-1.7$.  Within the uncertainties, the metal-to-metal ratios are consistent with the accretion of an H$_2$O-rich and rocky parent body, an 
interpretation supported by the anomalously high trace hydrogen.  The mixed atmosphere yields unusually short diffusion timescales for a helium
atmosphere white dwarf, of no more than a few hundred yr, and equivalent to those in a much cooler, hydrogen-rich star.  The overall heavy element 
abundances of the disrupted parent body deviate modestly from a bulk Earth pattern, and suggest the deposition of some core-like material.  The
total inferred accretion rate is $4.2\times10^9$\,g\,s$^{-1}$, and at least 4 times higher than any white dwarf with a comparable diffusion timescale.  
Notably, when accretion is exhausted in this system, both metals and hydrogen will become undetectable within roughly 300\,Myr, thus supporting 
a scenario where the trace hydrogen is related to the ongoing accretion of planetary debris.

\end{abstract}

\begin{keywords}
	circumstellar matter---
	stars: abundances---
	stars: individual (WD\,1536$+$520)---
	planetary systems---
	white dwarfs
\end{keywords}

\section{INTRODUCTION}

A decade of observational and theoretical studies by many astronomers has shown that, over a wide range of effective stellar temperatures, 
the presence of heavy elements in white dwarf atmospheres is evidence for orbiting planetary systems \citep{far16,van15,jur14,ver15}.  With
this relatively recent shift in paradigm, the discovery of the prototype, metal-lined white dwarf by \citet{van17} nearly a century ago -- while not 
a planet detection itself, but the signature of accreted planetary debris -- is arguably the first astronomical evidence of the presence of planetary 
systems around other stars \citep{zuc15}.

According to all dynamical models that deliver sufficient planetesimal masses into the innermost system where it can be accreted, each exoplanetary 
system hosted by a metal-enriched white dwarf must harbor at least a belt of minor bodies and one major planet \citep{fre14,ver13,deb12,bon11}.  The 
gravitational field of the planet(s) can perturb the orbits of the planetesimals onto orbits passing near the white dwarf so that they are tidally disrupted.  
{\em Spitzer} and complementary ground-based observations have established a firm connection between the atmospheric heavy elements in white 
dwarfs and the presence of dust and gas within the tidal radius of the star \citep{far09,von07,jur07,gan06}.

\begin{figure*}
\includegraphics[width=178mm]{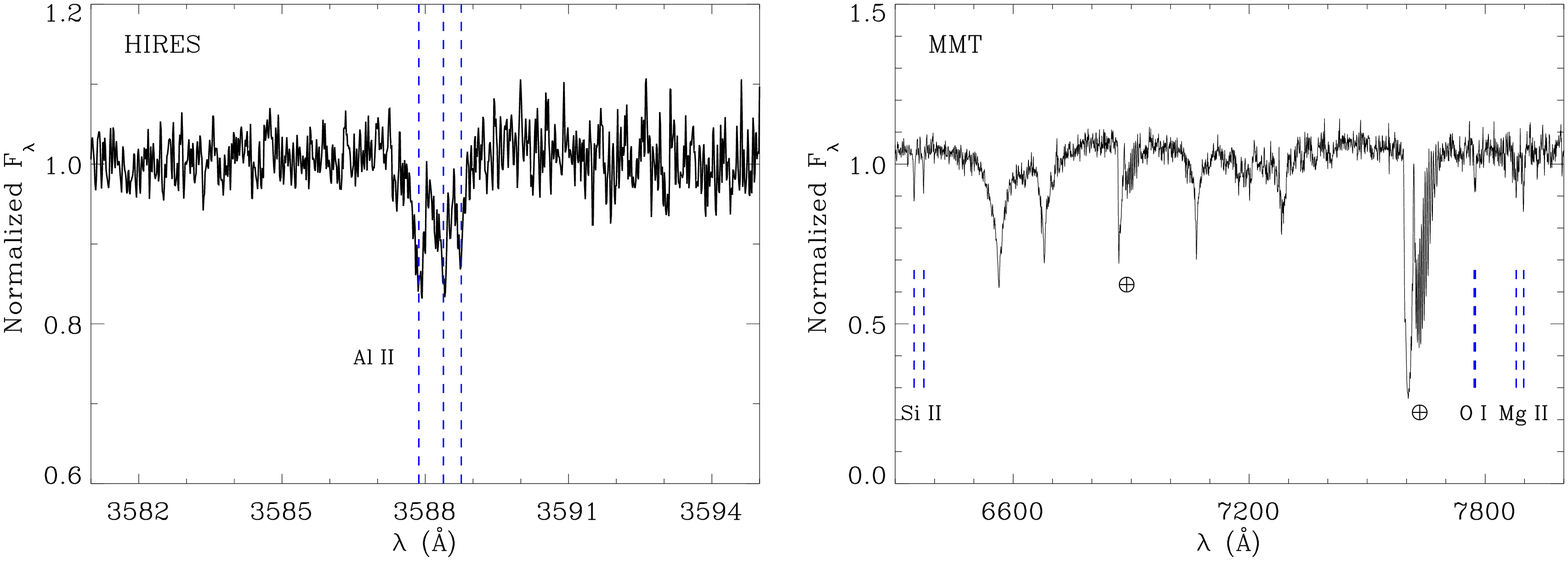}
\caption{In the left panel is a portion of the HIRES spectrum, with wavelength given in vacuum and showing a strong triplet of Al\,{\sc ii}.  The right 
panel shows a portion of the spectra obtained at the MMT, containing telluric features, H$\alpha$, two He\,{\sc i}, lines, and metal lines of O\,{\sc i}, 
Mg\,{\sc ii}, and Si\,{\sc ii}.  Unusually, the white dwarf exhibits strong lines of both H and He\,{\sc i}, indicating a mixed atmospheric composition.
\label{fig1}}
\end{figure*}

Because the metal-to-metal sinking timescales vary by no more than a factor of a few, the relative, steady-state abundances of the accreted 
planetary debris can be analytically linked to those observed in the polluted atmosphere \citep{koe09}, thus making the stellar surface an effective 
mirror of planetesimal composition.  The first detailed abundance study of any metal-enriched white dwarf was carried out for the current record 
holder for number (16) of detected heavy elements, GD\,362, demonstrating that the debris was broadly terrestrial-like \citep{zuc07}.  Since then, 
the broad pattern of bulk, Earth-like compositions has been seen -- especially with ultraviolet {\em HST} observations -- in several more stars with 
five or more heavy elements (O, Mg, Si, Ca, Fe) that indicate melting and differentiation among extrasolar, rocky planetesimals, and a diversity 
of overall compositions similar to different classes of Solar System meteorites \citep{xu13,gan12}.

Importantly, while most polluted white dwarfs appear to be contaminated by debris from parent bodies that were relatively poor in H$_2$O and 
other volatiles \citep{jur12a}, there is at least one case where substantial H$_2$O can be confirmed in an otherwise volatile- and carbon-poor 
planetesimal.  The debris orbiting and polluting the atmosphere of GD\,61 originated in a rocky minor planet roughly the size of Vesta and containing 
approximately 26\% water by mass \citep{far13b}.  Another polluted white dwarf with a substantial oxygen excess is SDSS\,J124231.07+522626.6,
where the parent body likely had an even higher water content \citep{rad15}.  Such water-rich asteroids are important as potential building 
blocks of habitable planetary surfaces, especially if most small and rocky planets form dry as did the Earth \citep{mor00}.  

This paper reports the identification and analysis of H, O, Mg, Al, Si, Ca, Ti, Cr, and Fe in the helium atmosphere white dwarf WD\,1536$+$520.  These 
elements are found to be accreting at a rate higher than any yet measured in a white dwarf with relatively short sinking timescales, and producing 
atmospheric metal abundances comparable to those of the Sun.  The data are consistent with a refractory-rich parent body with a modest fraction 
of H$_2$O.  Section 2 presents spectroscopic observations from several facilities that resulted in the detection of all the major rock forming elements, 
and strong upper limits on key volatiles.  The atmospheric modeling is discussed in Section 3, along with the determination of stellar parameters, and 
elemental abundances within the star and the disrupted parent body.  The paper explores the so-far unique properties of this star as something 
of a transition object between helium- and hydrogen-rich, polluted white dwarfs, with the conclusions presented in Section 4.

\section{OBSERVATIONS}

WD\,1536$+$520 was first identified in the Second Byurakan Sky Survey (SBS\,1536$+$520; \citealt{ste99,bal97}) in 1992 and correctly typed 
as a DBA (strongest lines He\,{\sc i}, weaker lines of H) white dwarf from a low resolution, $R\approx400$ spectrum.  It was spectroscopically 
observed as part the Sloan Digital Sky Survey in 2002 (SDSS\,153725.71$+$515126.9; \citealt{eis06}), and exhibits lines of Mg, Si, and Ca in 
these $R\approx2000$ data \citep{gan16}, yielding a full spectral type of DBAZ.  Given that the SDSS $ugriz$ photometry alone results in a 
temperature estimate of 22\,000\,K \citep{gir11}, the presence of these metal absorption features in a modest resolution spectrum is remarkable 
-- at similar $T_{\rm eff}$ and irrespective of atmospheric composition, the detection of atmospheric metals in white dwarfs typically requires 
powerful, high-resolution spectroscopy with Keck or the VLT \citep{koe05}.  The star has an infrared excess detected by {\em WISE} \citep{deb11,
bar14} at 3.4 and 4.6\,$\mu$m, where the data are consistent with passively heated debris orbiting within the Roche limit, similar to roughly 40 
other metal-enriched white dwarfs accreting from analogous disks \citep{far16}.

Follow up observations were obtained in 2014 April with the MMT using the Blue Channel Spectrograph.  Spectra were taken through a $1''$ 
slit with the 832\,l\,mm$^{-1}$ grating in first and second order, covering $6200-8100$\,\AA \ at 2\,\AA \ resolution and $3200-4100$\,\AA \ at 
1\,\AA \ resolution respectively.  The red spectrum consisted of four 900\,s exposures in clear conditions, while the blue spectrum comprised 
three 600\,s exposures but intruding on twilight where significant sky signal was present.  The blue data are thus of relatively modest quality, 
while the red spectra are superior and a combined spectrum is shown in Figure \ref{fig1}.  Most important, these modest resolution data 
exhibit a strong O\,{\sc i} 7775\,\AA \ absorption feature, in addition to lines of Mg\,{\sc ii}, and Si\,{\sc ii}. 

Additional, medium-resolution spectra were taken in 2014 July using the double arm ISIS spectrograph on the WHT.  Simultaneous blue 
and red spectra were taken through a $1''$ slit using the R1200B and R1200R gratings, with the 5300\,\AA \ dichroic, resulting in two spectra 
covering $4500-6000$\,\AA \ at a resolution of roughly 1\,\AA.  The white dwarf was observed continuously for eight exposures of 900\,s in 
good conditions.  The ISIS spectra reveal weak Al\,{\sc ii} and Si\,{\sc ii} features in wavelength regions not covered by the MMT dataset. 

Lastly, high-resolution observations carried out in 2015 April with the HIRESb spectrograph on Keck I.  The setup was identical to that 
described in \citep{zuc11}, covering the range 3130--5940\,\AA.  The blue cross disperser was combined with a $1\farcs15$ slit resulting in 
a spectral resolving power of $R\approx40\,000$.  Reduction procedures utilized both {\sc iraf} and {\sc makee}.  This dataset reveals multiple
lines of Mg\,{\sc ii}, Al\,{\sc ii}, Si\,{\sc ii}, Ca\,{\sc ii}, Ti\,{\sc ii}, Cr\,{\sc ii}, and Fe\,{\sc ii}, a portion of which is shown in Figure \ref{fig1}.

All spectra were reduced in the standard fashion, by average-combining each spectrum after extraction, using variance weighting for sky 
subtraction and rejection of bad pixels and cosmic rays.

\begin{table}
\begin{center}	
\caption{Stellar Parameters for WD\,1536$+$520\label{tbl1}}
\begin{tabular}{@{}lr@{}}

\hline

SpT								&DBAZ\\
$g$ (AB\,mag)						&17.06\\
$d$ (pc)							&$217\pm15$\\
$T_{\rm eff}$ (K)					&$20\,800\pm800$\\
$\log\,g$ (cm\,s$^{-2}$)				&$7.96\pm0.10$\\
Mass ($M_{\odot}$)					&$0.58\pm0.05$\\
$\log$\,(H/He)						&$-1.7\pm0.1$\\
$\log\,(M_{\rm cvz}/M)$				&$-11.16$\\
Cooling Age (Myr)					&$62^{+16}_{-6}$\\

\hline

\end{tabular}
\end{center}
\end{table}

\begin{table}
\begin{center}	
\caption{Lines Used for Abundance Determinations and Upper Limits\label{tbl2}}
\begin{tabular}{@{}ll@{}}

\hline

Ion			&Vacuum Wavelength (\AA)\\

\hline

C\,{\sc ii}		&4268\\
O\,{\sc i}		&7775\\
Mg\,{\sc ii}		&4435,4482\\
Mg\,{\sc i}		&5185\\
Al\,{\sc ii}		&3588,4664\\
Si\,{\sc ii}		&3854,3857,3863,4129,4132,5042,5057,6348,6373\\
P\,{\sc ii}		&3509,3787\\
S\,{\sc ii}  		&4163,5202,5214,5455\\
Ca\,{\sc ii} 	&3159,3737,3934\\
Ti\,{\sc ii} 		&3235,3237,3239,3342,3349,3350,3362,3373,3384,3686,3762\\
Cr\,{\sc ii}		&3336,3340,3343,3369,3404,3409\\
Fe\,{\sc ii} 		&3155,3163,3168,3178,3184,3194,3196,3211,3214,3228,3259\\
			&3290,3324,3469,3494,5170,5199,5200,5236,5277,5318\\

\hline
         
\end{tabular}
\end{center}
\end{table}

\section{ATMOSPHERIC PARAMETERS AND ABUNDANCE PATTERNS}

The multiple spectral datasets were analyzed together using white dwarf atmospheric models, where the input physics is detailed in \citet{koe10}.
The final stellar parameters were based on spectral fits to the latest SDSS spectrum, obtained with the BOSS spectrograph.  We calculated a 
3-dimensional model grid in $T_{\rm eff}$, $\log\,g$, and [H/He], keeping the latter fixed while fitting the first two parameters.  This is a more stable 
procedure than fitting for all three parameters, since the effect of [He/H] and $\log\,g$ on the spectrum is much smaller than that of the temperature. 
The results indicate the best fit is near 20\,800\,K, which was confirmed by repeating a similar fit with $\log\,g$ kept fixed and fitting for $T_{\rm eff}$ 
and [H/He].  The final stellar parameters are given in Table \ref{tbl1}.

For the determination of abundances and upper limits, all available spectra (Keck, MMT, SDSS, WHT) were used with the method of line profile fitting.  
Table \ref{tbl2} lists all the individual ions and wavelengths used for this purpose.  After a good fit was approximated, models were re-calculated with 
$\pm0.3$\,dex abundances, where visual inspection of each line determined the abundance and an error estimate.  In the case of multiple lines the 
final abundance was determined as a weighted average.  Repeating the analysis with $T_{\rm eff}$ and  $\log\,g$ varied within the adopted errors, 
the systematic errors were found to be approximately $0.20$\,dex, but in the same direction for all elements. Relative abundance errors are $0.05$\,dex.  
The changes of the convection zone and diffusion timescales contribute $0.10$\,dex; and thus the total systematic error in relative abundances is $0.12$\,dex.
 
\subsection{Bulk Composition of Accreted Debris}

The abundances, relative to helium, of all trace elements are given Table \ref{tbl3}, together with diffusion timescales for each species.  The 
third column compares the atmospheric, heavy element abundances in the white dwarf (relative to He) in units of solar values (relative to H;
\citealt{lod03}), demonstrating that WD\,1536 nominally exceeds the solar values for nearly all detected elements.  These absolute abundances 
surpass the previous record holder SDSS\,J073842.56+183509.6 \citep{duf10} by a factor of 3--10, and GD\,362 by over an order of magnitude 
\citep{xu13}.

\begin{table}
\begin{center}	
\caption{Abundances, Masses, and Accretion Rates for Trace Elements\label{tbl3}}
\begin{tabular}{@{}rrlrrr@{}}

\hline

			&			&[Z/He] 			&				&Early Phase				&Steady State\\
Element		&[Z/He]		&-- 				&$t_{\rm diff}$		&$X_{\rm z} M_{\rm cvz}$		&$\dot M_{\rm z}$\\
			&			&[Z/H]$_{\odot}$	&(yr)				&(10$^{19}$g)				&(10$^9$\,g\,s$^{-1}$)\\

\hline

H			&$-1.70$		&				&$\infty$			&4.012					&\\
C			&$\leq-4.2$	&				&261				&$\leq0.152$				&$\leq0.183$\\
O			&$-3.40$		&$-0.09$ 			&302				&1.276					&1.329\\
Mg			&$-4.06$		&$+0.39$ 			&166				&0.424					&0.804\\
Al			&$-5.38$		&$+0.16$			&141				&0.023					&0.050\\
Si			&$-4.32$		&$+0.14$			&122				&0.269					&0.692\\
P			&$\leq-7.1$	&				&120				&$\leq0.001$				&$\leq0.001$\\
S			&$\leq-5.4$	&				&126				&$\leq0.026$				&$\leq0.064$\\
Ca			&$-5.28$		&$ +0.38$ 		&146				&0.042					&0.091\\
Ti			&$-6.84$		&$+0.24$			&126				&0.001					&0.003\\
Cr			&$-5.93$		&$+0.43$			&103				&0.012					&0.037\\
Fe			&$-4.50$		&$+0.03$			&97				&0.354					&1.148\\

\hline
$\Sigma$		&			&				&				&2.40					&4.16\\

\hline

\end{tabular}
\end{center}

{\em Note}. Errors in abundance determinations are typically $0.1-0.2$ dex.  The fifth column is the mass of each element residing in the stellar 
convection zone, which consists of $8.0\times10^{21}$\,g of helium and $4.0\times10^{19}$\,g of hydrogen.  Due to their continual sinking, the 
mass of heavy elements within the convection zone represents a minimum mass for the parent body.  The metal-to-metal ratios within the 
planetary debris for the early phase and steady state regimes are derived directly from the values in the fifth and sixth columns respectively.
The diffusion timescales are a sensitive function of $T_{\rm eff}$ within the range of acceptable temperatures for WD\,1536, and thus
contribute some additional uncertainty to the derived abundance ratios.

\end{table}

Also calculated are the mass of each element present in the photosphere of the star, which is equivalent to the mass fraction of a given element 
$X_{\rm z}$, multiplied by the mass of the convection zone $M_{\rm cvz}$.  If WD\,1536 is in an early phase of accretion, where less than a single 
diffusion timescale has expired since the onset of atmospheric pollution, then the metal-to-metal abundances of the infalling debris are exactly 
mirrored by those in the atmosphere and given in the fifth column.  If instead pollution has been ongoing for at least 5 diffusion timescales 
\citep{koe09}, then the system is in a steady-state balance between accretion and diffusion and the abundance ratios are reflected in the sixth 
column.  For all detected elements but oxygen, the metal-to-metal ratios show little variation between the early phase and steady state solutions.

\begin{figure}
\centering
\includegraphics[width=84mm]{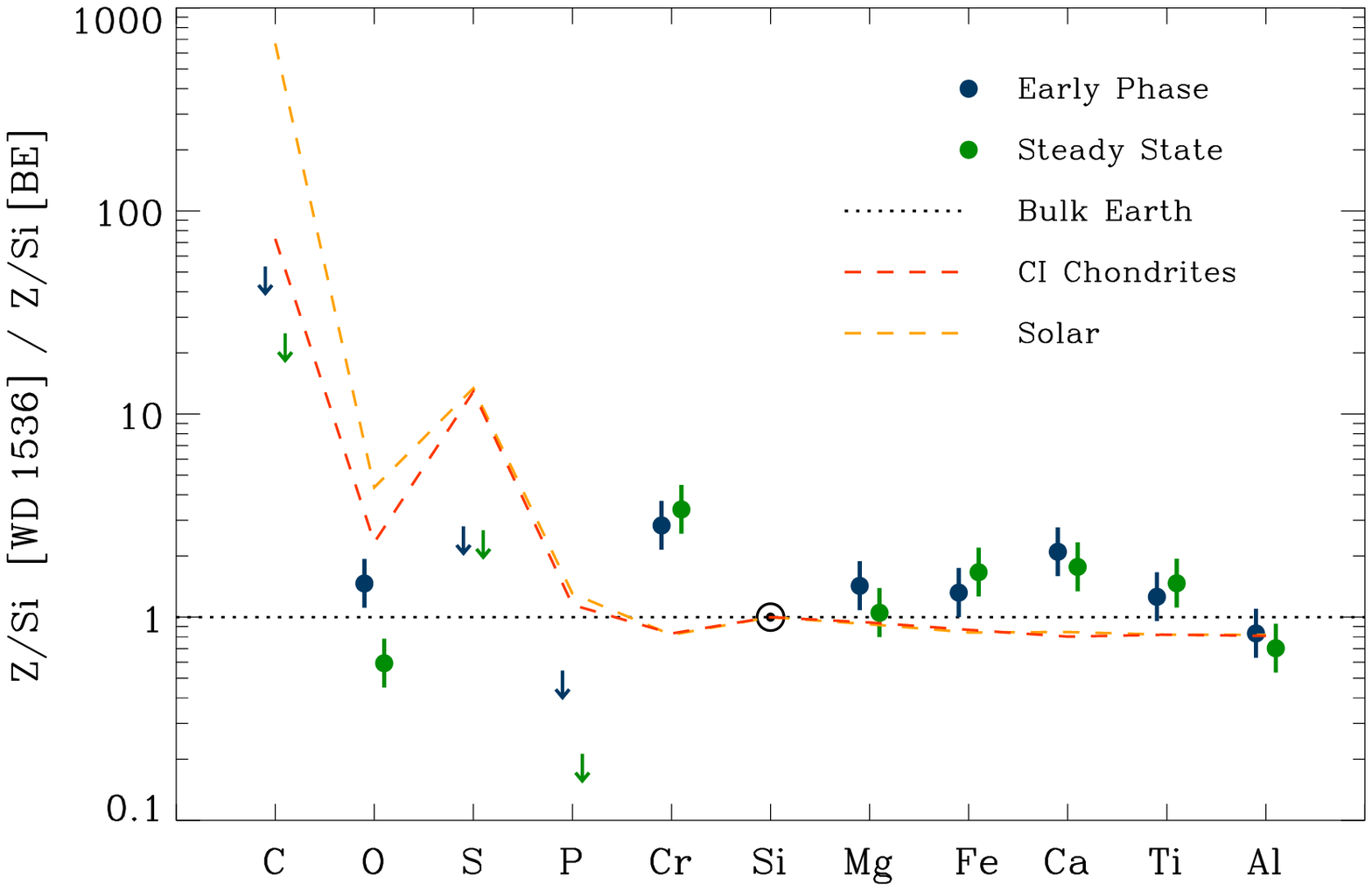}
\caption{Derived number abundances for the planetary debris polluting the outer layers of WD\,1536.  Shown in blue and green respectively, 
are the early phase and steady state abundances of each heavy element relative to Si, divided by the same ratio as determined for the bulk Earth 
\citep{all01}. A typical uncertainty in the derived ratios is 0.12\,dex, and error bars of this size are overplotted.  Also plotted are the same ratios for 
CI chondrites and the Sun \citep{lod03}, demonstrating that these compositions can be confidently ruled out for WD\,1536 based on ground-based 
data alone.  At face value, the disrupted parent body appears broadly similar to the bulk Earth, with notably high chromium.  
\label{fig2}}
\end{figure}

In Figure \ref{fig2} are plotted both the early phase and steady state abundances of heavy elements, relative to silicon and normalized to the 
bulk Earth values from \citet{all01}.  As can be seen, the debris orbiting and polluting WD\,1536 is bulk Earth-like in the major rock forming 
elements to within a factor of around two.  There is a notable enhancement in chromium, yet an apparent deficit in phosphorous.  This two-fold 
deviation in opposite directions is difficult to reconcile, as both chromium and phosphorous are siderophiles with similar condensation temperatures 
\citep{lod03}.  Similar enhancements in chromium have been seen in the white dwarfs PG\,0843+516 and GALEX\,J193156.8+011745  \citep{gan12} 
-- together with bulk Earth or higher phosphorous abundances, as expected -- but are otherwise not commonly seen in polluted white dwarfs 
\citep{xu14,jur12b}. Because phosphorous has only been detected in white dwarfs at ultraviolet wavelengths, the upper limit derived for WD\,1536 
from optical data may be uncertain.  With this caveat, the data are consistent with the accretion of substantial core-like material.

\subsection{Oxygen Excess and Hydrogen Accreted from H$_2$O}

The total oxygen budget can be evaluated by accounting for all the expected oxides originating in planetary solids \citep{far13b,kle10}.  
In the early phase and steady state scenarios, oxygen is first assumed to be carried exclusively by MgO, Al$_2$O$_3$, SiO$_2$, CaO, 
and FeO within the debris.  There are three possible outcomes from such an analysis.

\begin{enumerate}

\item{Insufficient oxygen to account for metal oxides.  This outcome can imply that iron was delivered not as FeO but substantially as metal.}

\item{An oxygen budget as expected for oxides in planetary solids.  In this case the debris is rocky and poor both in water ices and hydrated 
minerals resulting from aqueous alteration.}

\item{Excess oxygen beyond that of anhydrous minerals alone.  In this case, H$_2$O is the most likely source of the oxygen surplus.}

\end{enumerate}

Carbon can confidently be ignored as an oxygen carrier for the following reasons.  First, carbon has been found to be significantly depleted relative 
to solar and volatile-rich, cometary abundance patterns in nearly all polluted white dwarfs where measurements are available \citep{wil16,koe14,
far13a, jur12b,gan12,jur06}.  Second, CO and CO$_2$ are no more than 5\%--10\% of the volatile content of Solar System comets, which are 
dominated by water ice \citep{bin00}.  Third, for WD\,1536 in particular, the upper limit carbon abundance suggests that it cannot be a significant 
source of excess oxygen.

Table \ref{tbl4} evaluates the nominal oxygen budget for WD\,1536 for both an early phase and steady state accretion history.  In the steady 
state, there is insufficient oxygen to account for Mg, Al, Si, Ca, and Fe bound in oxides -- {\em only if 100\% of the iron was delivered as metal
or alloy can the oxygen budget be considered balanced and physical}.  In this case, the nominal oxygen abundance still requires a modest, 
$5-10$\% increase to account for the other elements (Mg, Al, Si, Ca) that only form rocks, but such leeway is well within the uncertainties.  
This is another strong indication that the material orbiting and polluting the white dwarf has a substantial core-like component.  Of the total iron 
mass present in the Earth, metallic Fe in the core is thought to represent 87\%, whereas Fe in the mantle and crust is only 13\% \citep{mcd00}, 
some of which is also metal.  Thus, the scenario where the iron in WD\,1536 was contained essentially in pure metals or alloys is plausible.  
Within the derived photospheric abundance errors, a steady state solution without any iron oxides would readily allow for solutions where the 
parent body contained water ice or hydrated minerals.

\begin{table}
\caption{Oxide, Iron Metal, and Water Mass Fractions\label{tbl4}}
\begin{center}	
\begin{tabular}{lrr}

\hline

Oxygen Carrier		&Early Phase		&Steady State\\

\hline

MgO				&0.22			&0.40\\
Al$_2$O$_3$		&0.02			&0.03\\
SiO$_2$			&0.24			&0.59\\
CaO				&0.01			&0.03\\
FeO$^a$			&0.08			&0.00\\

\hline
O Excess:			&0.43			&--\\
H$_2$O in debris:	&0.25			&--\\

\hline
Fe in metal:		&0.00			&1.00\\

\hline

\end{tabular}	
\end{center}

$^a$ Upper limit for FeO.

\smallskip
{\em Note}.  The first five rows assumes oxygen is carried to maximum capacity by all the major rock forming elements, but in fact iron can also 
be in pure metal or iron-nickel alloy with no oxygen.  The nominal oxygen budget in the steady state is unphysical unless 100\% of the total iron is 
carried as metal, and the nominal O/Si and O/Mg ratios are marginally higher than tabulated.

\end{table}

While the range of allowed abundance ratios also permits solutions without any excess oxygen, the striking hydrogen abundance in WD\,1536 
must be considered, and which clearly favors a water-rich interpretation.  While an early phase of accretion predicts an oxygen excess and thus 
the need for H$_2$O within the planetary debris, the heavy element settling times are relatively short, and thus catching the star in this phase
is less likely.  If disks last for at least 10$^5$\,yr \citep{gir12}, then the probability that WD\,1536 is not yet in a steady-state phase of accretion 
is less than 1\%.  The total hydrogen mass within the stellar atmosphere is $4.0\times10^{19}$\,g, and could have been delivered by an asteroid 
with total mass a few to several times 10$^{21}$\,g and which was 5--10\% H$_2$O by mass.  This would be consistent with the lower mass limit 
of $4.2\times10^{19}$\,g from the heavy elements alone.

While uncertain, the totality of data discussed in this section favors the deposition of H$_2$O onto the stellar surface and carried by the parent 
body whose debris now orbits the star.  In the next section, the anomalously high trace hydrogen abundance is shown to be transient, thus 
strengthening this interpretation.

\subsection{Anomalous Diffusion Timescales and Trace H}

The mass of the convection zone in WD\,1536 is tiny -- $10^6$ times smaller than those within the bulk of known polluted white dwarfs with 
helium atmospheres.  There are two reasons for this.  First, the $T_{\rm eff}$ and 60\,Myr cooling age mean the star is experiencing the early 
stages of convection zone growth \citep{paq86}.  In fact, with $T_{\rm eff}>20\,000$\,K this star is the warmest and youngest helium-rich white
dwarf to show metals due to ongoing accretion.  Second, the anomalously high fraction of hydrogen leads to a significant reduction in the depth 
of the outer layers relative to a pure helium composition, by a factor of approximately 30.  

\begin{figure}
\centering
\includegraphics[width=84mm]{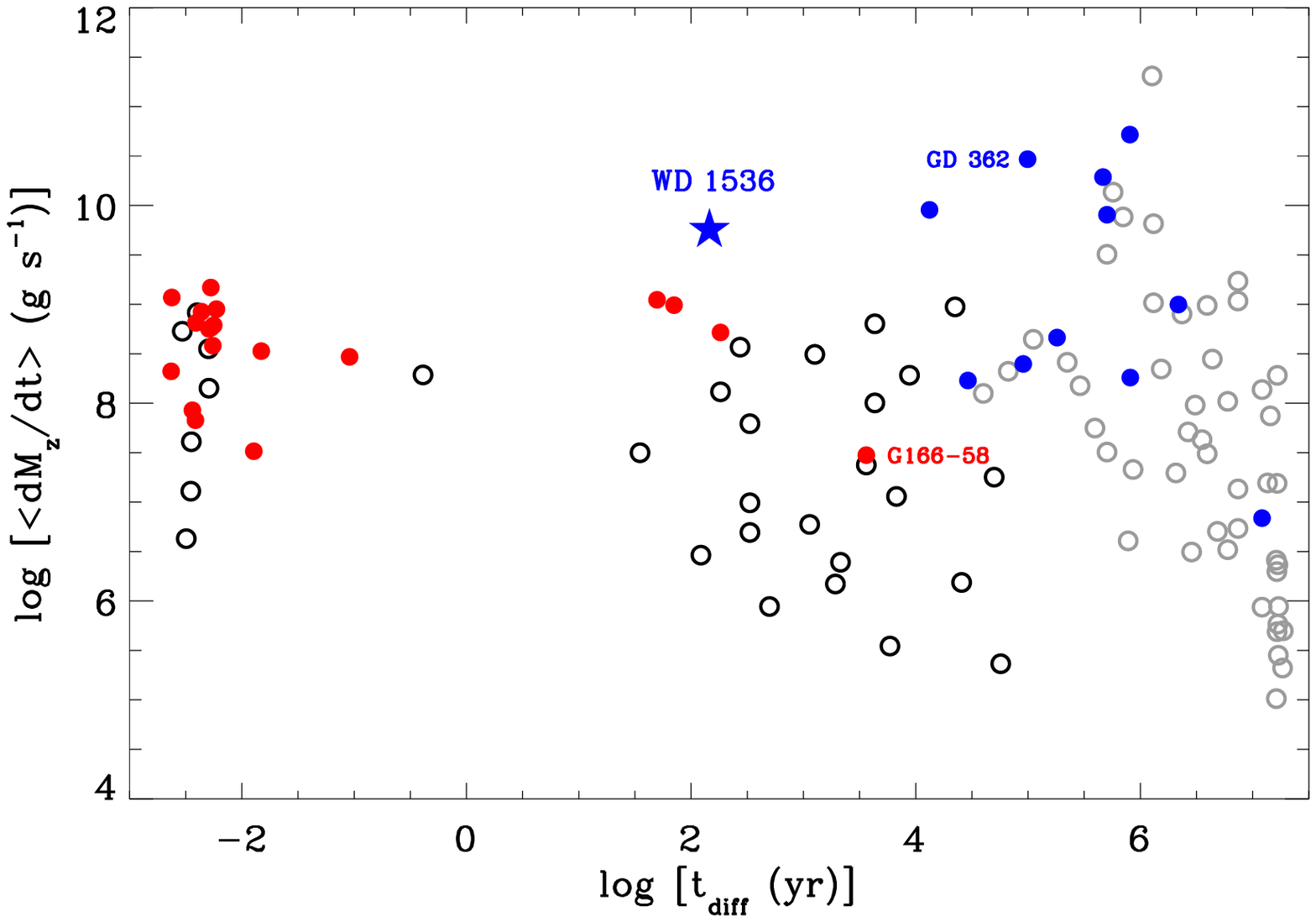}
\caption{Accretion rate versus diffusion timescale for WD\,1536 and a large sample of metal-enriched white dwarfs observed with {\em Spitzer} 
\citep{ber14}.  For consistency, all plotted rates and timescales are based solely on Ca, following the method outlined in \citet{far12} with updated 
diffusion data ({http://www1.astrophysik.uni-kiel.de/{\raise.17ex\hbox{$\scriptstyle\sim$}}koester}; \citealt{koe09}).  The hydrogen-rich stars are shown 
as red filled and black open circles, while the helium-rich stars are shown as blue filled and grey open circles; filled symbols correspond to the detection 
of infrared excess.  Within each atmospheric class, left to right represents decreasing $T_{\rm eff}$.  Remarkably, WD\,1536 sits in a region that is 
otherwise exclusively occupied by stars with hydrogen atmospheres.  G166-58 is the coolest white dwarf with a hydrogen-rich atmosphere and an 
infrared excess.
\label{fig3}}
\end{figure}

\begin{figure}
\centering
\includegraphics[width=84mm]{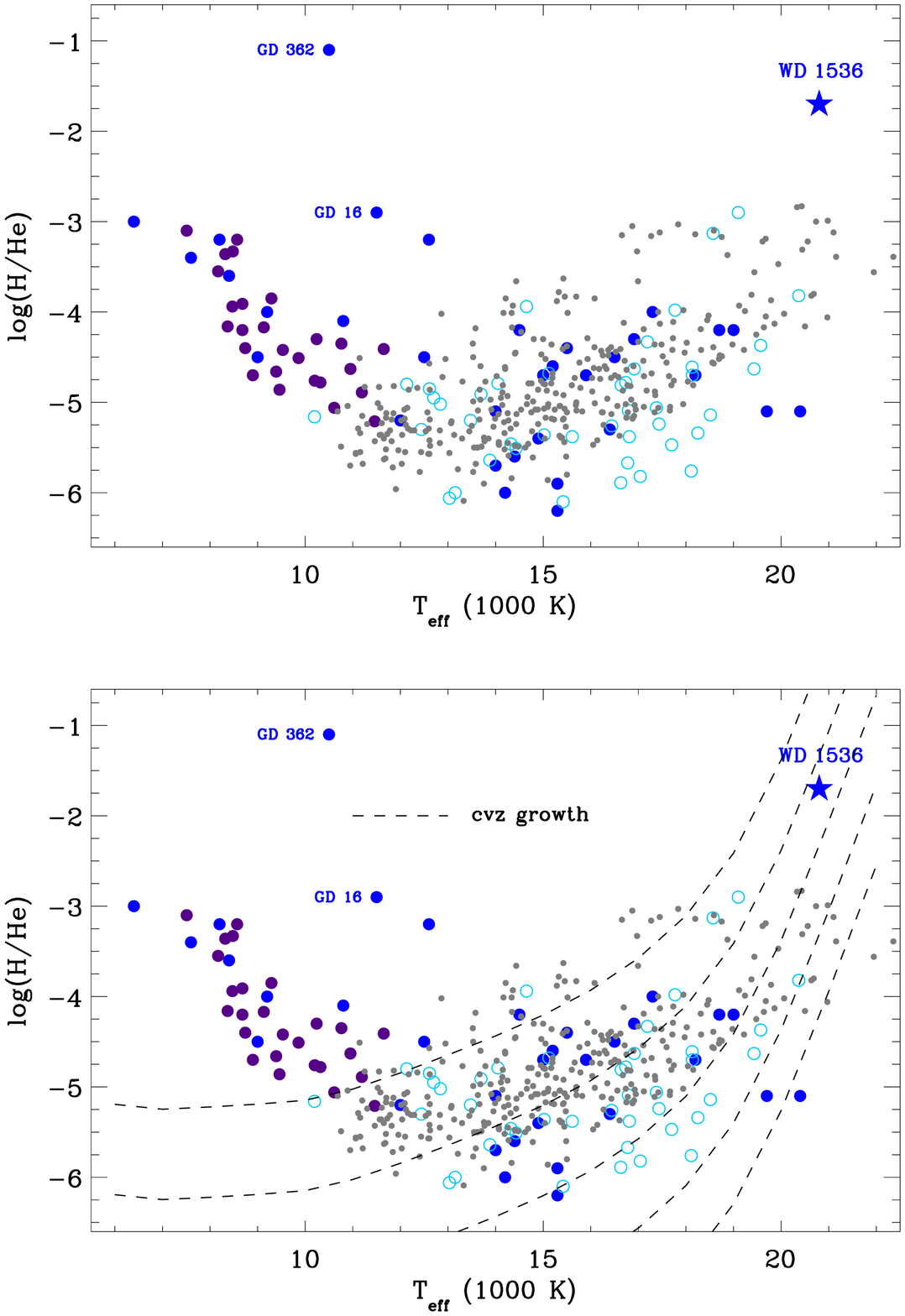}
\caption{Relatively cool, helium atmosphere white dwarfs with detected hydrogen absorption.  Light blue open circles are nearby stars from \citet{ber11} 
and \citet{vos07}, while the grey dots are a more distant and abundant sample from \citet{koe15}.  Filled purple circles are stars with strong Ca\,{\sc ii}
from \citet{duf07}, and filled dark blue circles are Figure \ref{fig3} stars with matching criteria.  The upper panel demonstrates the anomalous nature of 
WD\,1536, whose H/He ratio is over an order of magnitude higher than any comparable object and second only to GD\,362 \citep{zuc07}.  Overplotted 
in the lower panel are tracks demonstrating the directly proportional change in H/He due to the growth of the convection zone over time in pure helium 
atmosphere stars \citep{koe09}, shown for five orders of magnitude of initial [H/He] $=1.0,0,-1.0,-2.0,-3.0$.  These calculations assume complete mixing 
of the outer layers.
\label{fig4}}
\end{figure}

These facts conspire to make the stellar atmosphere physically similar to a typical 10\,000\,K hydrogen-rich white dwarf, with comparable diffusion 
timescales.  Figure \ref{fig3} plots WD\,1536 together with a sample of polluted white dwarfs observed with {\em Spitzer}, as a function of their 
inferred accretion rates and sinking timescales based on Ca\,{\sc ii} detections \citep{ber14}.  WD\,1536 lies above three hydrogen-rich stars 
whose disks have been detected in the infrared \citep{far10b,von07}, and has an even shorter sinking timescale than G166-58 \citep{far08}.
This so-far unique position for a helium-dominated star strongly suggests 1) it is accreting at a high rate in a steady state, and 2) that the older, 
cooler stars with similar sinking timescales do not often experience similarly high rates of accretion.  If this interpretation is correct, it would support
a decreasing trend of planetary dynamical activity in the post-main sequence, as measured by the mass influx of planetesimals towards the white 
dwarf host, consistent with theoretical predictions \citep{mus14,ver13}

Figure \ref{fig4} highlights the exceptional H/He in WD\,1536.  The upper panel plots samples of helium-rich white dwarfs with trace hydrogen 
detected directly through Balmer absorption features (typically only H$\alpha$)\footnote{WD\,1536 displays strong Balmer lines up to and including 
H$\delta$.}.  Interestingly, a substantial fraction of the plotted stars are also polluted with heavy elements, although a strong bias is present at 
$T_{\rm eff}\la12\,000$\,K.  In this cooler temperature range, He\,{\sc i} absorption rapidly becomes too weak to detect in low- and medium-resolution 
spectra, whereas strong Ca\,{\sc ii} absorption can indicate a helium-rich atmosphere \citep{duf07}. At the warmer end of the temperatures shown in 
Figure \ref{fig4}, the bias towards metal detection is not an issue.  Caution should be used when viewing Figure \ref{fig4}; the plotted stars do not 
represent an evolutionary sequence, and selection biases play a large role.  That being said, the cooler stars with substantial hydrogen are either
born with substantially more massive reservoirs than can currently be inferred in earlier evolutionary stages, or accrete H-rich planetary material.

\subsection{H/He Evolution}

Assuming complete mixing of the outer stellar layers, the lower panel of Fig \ref{fig4} plots tracks of constant hydrogen mass within otherwise-pure 
helium atmosphere white dwarfs, as a function of temperature \citep{koe09}.  In this simple model where no stratification occurs between hydrogen 
and helium, the observational signature of most fixed masses of hydrogen at $T_{\rm eff}\approx20\,000$\,K will gradually disappear from white
dwarfs with helium-dominated atmospheres.  The fact that some stars retain (or re-gain) substantial hydrogen masses at later times, is a well-known 
problem in white dwarf atmospheric evolution \citep{mac91}.  While this general topic is beyond the scope of this paper, two distinct possibilities are 
1) hydrogen is accreted over long timescales, or 2) primordial hydrogen floats over a deeper helium reservoir, and is later mixed into the photosphere.  
In the latter scenario, stars will appear hydrogen-rich at sufficiently warm temperatures and later reveal themselves to be helium-dominated \citep{ber11,
fon87}.  Currently, there is more observational support for the primordial model, with at least 3/4 of helium atmosphere white dwarfs showing traces of 
hydrogen \citep{koe15}.

{\em Thus in the absence of continued accretion WD\,1536 will have both its metals and trace hydrogen wiped clean from its photosphere}.  Without 
the influence of ongoing, external pollution, the heavy elements will completely sink beneath the photosphere within a few 10$^3$ years at most.  But 
over longer timescales, the remarkably high abundance of atmospheric hydrogen will be drowned by the deepening helium convection zone.  By the 
time WD\,1536 has cooled to 15\,000\,K in 140\,Myr, the mass of the convection zone will have grown by 4 orders of magnitude and exhibit a trace 
hydrogen abundance $\log$(H/He) $=-5.7$.  At this stage, the star will either appear as a fairly average helium-rich white dwarf, where hydrogen is 
difficult or impossible to detect in modest resolution spectra due to its apparent faintness relative to the nearby samples shown in Fig \ref{fig4}.  When 
the star has cooled to 12\,000\,K after another 190\,Myr, it will certainly not have detectable hydrogen.  The unavoidable conclusion is that this white 
dwarf is being witnessed at a special time, in a transient phase, and the hydrogen is related to the orbiting planetary debris, and thus water is likely 
present.

Considering that WD\,1536 may have only accreted $\sim10^{19}$\,g of hydrogen onto its atmosphere and $\log$(H/He) $=-1.7$, it can be seen that 
if a significantly larger and water-rich parent body had been deposited, then hydrogen could (temporarily) have become the dominant atmospheric 
constituent.  For example, the disrupted asteroids polluting GD\,61 or SDSS\,1242 might have delivered $\sim10^{21}$\,g or $\sim10^{23}$\,g of 
hydrogen respectively, resulting in abundances of $\log$(H/He) $\sim0$ and $\log$(H/He) $\sim+2$.  In both cases WD\,1536 would temporarily 
appear as a hydrogen-rich star despite being dominated by the underlying helium.  Therefore, the accretion of water-rich planetary debris has the
potential to have an observable effect on H/He white dwarf spectral evolution.

\section{CONCLUSION}

The young, helium-atmosphere, white dwarf WD\,1536 exhibits the highest abundances of heavy elements yet seen among polluted hosts of evolved 
planetary systems.  In addition to the broadly solar abundances of the major rock forming elements O, Mg, Al, Si, Ca, and Fe, this star also has a 
remarkably high trace hydrogen abundance of $\log$(H/He) $=-1.7$.  Considering the 1) abundance pattern of heavy elements, 2) the anomalously high 
trace hydrogen, and 3) the transient detectability of both the metals and the hydrogen, the most realistic conclusion is that the parent body whose debris is 
both orbiting and polluting WD\,1536 contained traces of H$_2$O.

The thinness of the convection zone is a result of relative youth and relatively high mass of trace hydrogen within a helium-dominated atmosphere. Due 
to these combined facts, the outer layers of WD\,1536 essentially behave as a hydrogen-rich white dwarf, with metal sinking timescales of only a few 
hundred years at most, hence supporting a steady state interpretation of the metal abundances.  If these are indeed in a steady state, then WD\,1536 
has the highest {\em instantaneous} accretion rate yet observed among polluted white dwarfs.

\section*{ACKNOWLEDGMENTS}

J. Farihi thanks S. Desch for feedback on a draft manuscript.  The authors acknowledge both the MMT and WHT (Service program SW2014a39) for 
the expedient use of their Directors' time, without which these results would not have been possible, and an anonymous reviewer for feedback that
improved the quality of the manuscript.  J. Farihi gratefully acknowledges the support of the STFC via an Ernest Rutherford Fellowship.  This research 
was supported in part by a NASA grant to UCLA, and by an NSF pre-doctoral fellowship to L. Vican.  The research leading to these results has received 
funding from the ERC under the European Union's $7^{\rm th}$ Framework Programme n.\ 320964 (WDTracer).

\label{lastpage}

\end{document}